# Molecular recognition of the environment and mechanisms of the origin of species in quantum-like modeling of evolution

**Alexey V. Melkikh,**

Ural Federal University, Yekaterinburg, 620002, Mira str. 19, Russia,

melkikh2008@rambler.ru

**Andrei Khrennikov,**

International Center for Mathematical Modelling in Physics and Cognitive Sciences, Linnaeus University, Växjö, S-35195, Sweden,

andrei.khrennikov@lnu.se

A review of the mechanisms of speciation is performed. The mechanisms of the evolution of species, taking into account the feedback of the state of the environment and mechanisms of the emergence of complexity, are considered. It is shown that these mechanisms, at the molecular level, cannot work steadily in terms of classical mechanics. Quantum mechanisms of changes in the genome, based on the long-range interaction potential between biologically important molecules, are proposed as one of possible explanation. Different variants of interactions of the organism and environment based on molecular recognition and leading to new species origins are considered. Experiments to verify the model are proposed. This bio-physical study is completed by the general operational model of based on quantum information theory. The latter is applied to model epigenetic evolution.

Keywords: speciation, molecular recognition, entanglement, quantum control, quantum information

1. **Introduction**

On the other hand, *the problem of origin of complexity, which is important in relation to living beings, is unsolved.* This problem has many aspects. For example, we can highlight the algorithmic complexity, computational complexity, information complexity and statistical processing of information. Different definitions of the complexity of living systems are considered in (Heylighen, 1999, Dawkins, 1986, Miconi, 2008, Piqueira, 2009, Finlay, Esteban, 2009, Marquet, 2000, Gell-Mann, 1994, Crutchfield, 2003, Salthe, 2008). In particular, two papers (Melkikh, 2014, 2015) emphasized that this problem (associated with the need to enumerate an exponentially large number of genomic variants) should be solved on the basis of precise mathematical formulation.

In regard to the evolution of life, we are always dealing with complex structures. This is one of the most important properties of living systems, without exception. Even some of the simplest single-celled organisms, such as archaebacteria, or cyanobacteria, have genomes that are approximately equal to the $10^6$ pairs of nucleotides. These genes encode complex systems of substance transport, information reception, energy conversion and many other processes.

Consider a chain of nucleotides of length N. There are $4^N$ variants of such sequences. How large is this number? For example, for N = 1000 we receive $4^{1000}=10^{602}$. Note here that N = 1000 corresponds to only one modern gene. For a genome size of $10^6$-$10^9$ the number of variants in any case is exponentially large and cannot be enumerated during the lifetime of the universe. In this sense, the problem of enumeration of genomic variants – the combinatorial problem - is the key to evolution. Without its solution it is impossible to speak about the adequacy of the theory.

A. Melkikh (2014) presented some reasons that such an algorithmic formulation of the problem leads to the following dilemma:

- Evolution is a priori undirected, but then it is impossible to prove a rational mechanism for the selection of variants of an exponentially large number. This applies to all mechanisms, including sexual reproduction, the selection of alleles in a population, and phenotypic plasticity.

- Or, evolution is a priori directed (i.e., it is known *a priori* that certain blocks encode something good). However, it is then difficult to justify the existence of such mechanisms in the framework of Darwinism. The essence of Darwinism is that a priori evolution will not focus elsewhere, it has no purpose, and species cannot know what they will need in the future. These are the axioms without which Darwinism does not exist.

To solve this problem, a mechanism for partially directed evolution was proposed (Melkikh, 2014, 2015, Melkikh, Khrennikov, 2016). The term "partially" reflects the fact that in any case, uncertainty in some form will be present in the environment, even if evolution was completely directed. This is due to the factors such as the uncertainty of the climate, and different random events such as asteroid strikes.

In particular, in the frame of the model of partially directed evolution it is possible to consistently explain many different evolutionary phenomena, such as the finite lifespan of organisms, the existence of the sexes, the genetic diversity of populations, the effect of the Red Queen, and phenotypic plasticity (Melkikh, Khrennikov, 2016).

Of course, the notion of ``directed evolution'' has to be formalized and this is a complex problem. For the moment, we use it heuristically and its essence is illustrated by important biological and bio-physical examples. The minimalist interpretation of ``directed evolution'' has the Lamarckian feature: changes in biological organisms are adapted to environment during organisms' life-time. (Thus not simply Darwinian mutations combined with post-selection generated by the environment.) This type of so-to-say instantaneously directed evolution is illustrated by the quantum-like model of epigenetic evolution proposed by Asano et al. (2013), see section 6 for the brief presentation of its basics. However, one of the coauthors, see Melkikh (2014), proceeds with a stronger interpretation. In his works it is presumed that biological systems (primary at the genetic level) can select the ``optimal evolutionary pathway'' and this optimization ``drive'' plays the active role in their evolution.

Among the major challenges to be answered by the theory of evolution are the following:

**Q1:** *How does biological complexity arise?*

**Q2:** *How do new biological systems arise?*

**Q3:** *How do new species cross the "ravines" in the fitness landscape?*

**Q4:** *Why are the molecular-genetic control systems stable?*

The article is devoted to the detailed mechanisms of partially directed evolution towards evolutionary innovation and speciation.

## 2. Biological complexity and evolution

From the mathematical point of view, we can identify algorithmic complexity, computational complexity, information complexity and statistical processing of information. In varying degrees, all these types of complexity may be relevant to the modeling of living systems. However, if the underlying problem is to consider the emergence of complex systems during the process of evolution, it seems that computational complexity is the most relevant. Computational complexity is associated with the characteristics of a mass of problems (as opposed to individual tasks of algorithmic complexity).

There are many classes of computational complexity (complexity of algorithms for computing), the most important of which are *P* and *NP*. The first is a polynomial algorithm in which the number of steps depends on a power of the number of elements of the system being analyzed. The class *NP* includes algorithms in which the number of steps depends exponentially on the number of elements. The question of reduction of *NP*-problems to *P*–problems is fundamental and has so far not been solved (see, for example, Aaronson, 2005).

*NP*-hard problems include, for example, the traveling salesman problem, the problem of satisfiability of a logical scheme and others. Polynomial algorithms (not enumerating) for such problems have not yet been found. *NP*-hard problems include some problems of game theory "against nature" in which aprioristic information about the opponent's moves is absent.

Piqueira (2009) reviewed various definitions of the complexity of living systems. According to this author, complexity is associated with the fact that a system consists of several parts that cannot be reduced to a simple summation of them.

Miconi (2008) also noted that there are various definitions of complexity. For example, complexity can be defined as the amount of information needed to describe an object using the Shannon entropy. There is also functional complexity. We intuitively associate the great complexity of organisms with a low probability of occurrence (see also Heylighen, 1999, Dawkins, 1986).

It was noted that adaptive functional complex systems are rare among all possible systems. What are the methods for the creation of such systems? It is argued that the random walk is not effective for this problem but that Darwin's evolution is effective. It is also argued that complex systems are easier to build from already existing complex systems. The author calls this assertion "Darwin's heuristics", which is "to look near the previously found". However, to implement such an algorithm, it is necessary to define the term "near". How does the organism decide whether it has strayed far enough from the "previously found"? To do this, it must have some prior information about the features space. The same can be said of the synthesis of complex systems from components.

Salthe (2008) noted that there is a positive definition of complexity (Gell-Mann, 1994, Crutchfield, 2003): ''effective'' or ''structural'' complexity, a concise listing of the regularities shown by a system (note that this definition actually coincides with algorithmic complexity). There is also a negative definition: in complex systems, situations arise that cause a surprise.

A number of articles (Finlay, Esteban, 2009, Marquet, 2000) discuss the correlation between complexity and various parameters of the organism and of populations (body mass, population size, trophic levels etc.).

Several authors believe that the increasing complexity of organisms is a natural consequence of Darwinian evolution. For example, Adami and co-authors (2000) noted the growth of complexity in the evolution of populations of artificial organisms. The definition of complexity based on the Shannon entropy was considered.

However, Davies (2004) noted that the increase in complexity can be explained by the fact that in the case of simple structures the system uses all of the allotted phase space that contains all of its allowable complex structure. Complex structures are numerous, so the movement toward sophistication is more likely. This is just one of the basic principles of probability theory. It is applicable not only to the microparticles, but also to arbitrary objects. What are the characteristic times of their appearance (especially in the context of computational complexity)? What are the conditions under which the complexity increases?

Jablonka and Lamb (2006) discuss the role of epigenetic mechanisms in the increase of complexity during evolution.

Schuster (1996) notes that it is impossible to understand what the difficulty is if we do not understand its origin. He examines several aspects of complexity, ecological diversity, internal complexity in the sense of logical depth and hierarchical complexity, and notes that large jumps in evolution are characterized by increasing complexity. On the basis of Darwinian dynamics - walking on the fitness landscape - he proposes a mechanism of complex systems origin. The mechanism consists of random walks that lead to small peaks on the landscape and the occasional shift of neutral networks, which in turn leads to higher peaks. It is noted that the principle of natural evolution consists of the construction of "new" forms based on previous versions.

The modular principle of genome design was proposed by Gilbert (1978). According to Schuster, block-hierarchical structures, which are typical of structures in wildlife populations, accelerate evolution. The role of hypercycles in the formation of more complex systems from replicators was discussed.

Summarizing the different approaches to the complexity of living systems, we can say that on the one hand we can ask questions about how to operate such a complex system (i.e., live), but the other question that arises in this case is "how in the process of evolution did such a system arise?". Obviously, these are related but different questions. Accordingly, for each of these different issues there may be suitable and different definitions of complexity.

One of the main drawbacks to consider regarding the emergence of complex living systems is that complexity is not usually considered mathematically; in fact, many definitions of complexity are intuitive. However, some properties of the systems are implicitly postulated. This does not allow us to reach unambiguous conclusions regarding the mechanisms of solutions to problems related to evolution. For example, it may be noted that not all problems are solved by an enumerating search based on block methods. Tasks such as breaking the password cannot be solved in this way in principle. Consequently, it is necessary to mathematically define the class of solvable problems of an evolutionary search as well as to determine the conditions under which the search takes place.

Thus, by considering the problems of biological complexity, we can draw the following conclusions:

- A priori undirected evolution does not necessarily lead to increased complexity. Heuristics such as "look near the existing" cannot be justified within the framework of undirected evolution.

- A priori directed evolution naturally leads to an increase in complexity. In this case, the problem of enumeration of exponentially large number of variants does not appear, and the *NP*-hard problem reduces to a *P*-hard problem. Exactly this case corresponds to heuristics of the type "search near existing."

### 3. Mechanisms of speciation and evolutionary innovations

Currently, most evolutionists agree that species origin takes place through natural selection. However, how exactly does selection lead to the formation of species? It is accepted in the literature (see, for example, Schluter, 2009) to allocate two large groups in the formation of species: ecological speciation and speciation associated with mutations. For example, in papers (Hubbs, 1940, Schluter, 2009), the evolution of parasitic fishes was considered. It concluded that because it is a repetitive process, it is caused by selection, but not by an accident. The author also noted that speciation is under environmental control. Experiments with *Drosophila* and yeast confirm the ecological mechanism, but the genetics of environmental speciation is currently poorly understood (Schluter, 2009).

On the other hand, mechanisms of speciation within the Darwinian theory of evolution have historically been associated with allopatric and sympatric species formation (see, Diekmann, Doebeli, 1999). Sympatric speciation refers to the formation of two or more descendant species from a single ancestral species, all occupying the same geographic location. During allopatric speciation, a population splits into two geographically isolated populations. Intermediate cases also exist. In peripatric speciation (sub-form of allopatric speciation) new species are formed in isolated, smaller populations that are prevented from exchanging genes with the main population. It is related to the concept of a "founder effect", since small populations often undergo bottlenecks. Genetic drift is often proposed to play a significant role in peripatric speciation. In parapatric speciation, there is only partial separation of the zones of two diverging populations, and individuals of each species may come in contact or cross habitats from time to time.

Mechanisms of species origin are naturally associated with the complexity of the fitness landscape (Gavrilets, 2010). The idea of the fitness landscape was developed by Wright (1932) and Simpson (1953). When the fitness landscape is highly fragmented, intersection of the valleys to reach higher peaks becomes an important problem. One way to solve this problem is genetic drift, which is important for small populations. Author (Gavrilets, 2010) notes that if there are 1,000 genes, each of which has only two alleles, the number of genomic variants will be:

$$2^{1000} \approx 10^{100}.$$

As a way out of this situation of enumeration of exponentially large numbers of variants, the author considers the main proposal of Maynard Smith, who considered the analogy with a game of words. In this case, only one letter of a word at each step may vary. If the correct letters are fixed, the given word can be achieved in a relatively short (polynomial) number of steps. The author proposed a two-dimensional grid as a model for evolution and examined percolation between different clusters. In this sense, a network of genotypes is largely similar to the lattice in liquid models.

The concept of evolvability was introduced by Dawkins (Dawkins, 1989) and is an important aspect of the modern theory of evolution. Dawkins suggested, based on the analysis of experimental facts, that in every generation animals must not only successfully survive but also more effectively evolve (for example, insects). He also suggested that there is a higher-order selection, which increases the ability of the organism to evolve. This very feature can also evolve. According to Dawkins, this is akin to the selection of clades (each clade represents a separate branch of the tree of life). Some branches have a high evolutionary potential. According to (Dawkins, 2009), this selection is different from Darwinian selection.

Janković (2016) believes that the laws of nature are not favorable for life. However, if life one way or another appears, with it appear mechanisms for its preservation. The author pays great attention to the evolvability concept, considering it one of the most important for the understanding of life. According to Janković evolvability can be defined as the ability to evolve in a changing environment:

*Evolvability of a biosphere is the measure of summary potential of evolutionary change of all its living beings, together with some measure of overall propensity of its systems to undergo evolutionary change upon given conditions.*

According to the author (Janković, 2016), evolutionary changes are random in the sense that they are not directed in advance to some purpose. However, they are not completely random, in the sense that they use the last structures. The author considers the example of the evolution of Darwin's finches, and believes that the paradox of time for them to be solved by the fact that the process of evolution is not just brute force, but a cooperative process, including the use of previously existing beaks. It is important, however, to determine the mechanisms of a "partially random" evolution.

One of the manifestations of the evolvability concept, according to (Janković, 2016), may be a change of the whole biosphere to increase the capacity of species to survive. This assumption is largely similar to the concept of Gaia, which was repeatedly discussed in the literature (Lovelock, 1979, McDonald-Gibson et al, 2008, Kleidon, 2010, Boyle et al, 2011, Chopra and Lineweaver, 2016).

The author believes that the information plays an active role in evolution and proposes to consider the information as a basis for the definition of life. The author has defined the coding concept for living systems as the sum of the different stable states of physical and chemical systems that can be used as the basis for the maintenance of genetic information, together with the rules governing the flow of such information. In the case of earthly life, such a framework comprises the nucleotides of DNA.

During evolution, the coding framework (the genetic code) has changed very little. According to the author, the larger the space of the parameters, the greater the potential for building the taxa.

We want to emphasize, however, that evolvability can be justified only in the frame of partially directed evolution (see, Melkikh and Khrennikov, 2016).

Information aspects of evolution were also discussed by other authors (see, e.g., Yokey, 2000, 2002, Trevors and Abel, 2004, D'Onofrio et al., 2012, Wills et al, 2015).

Wilson (2010) considered the problem of multi-level selection that is actively discussed now (Allen et al, 2013, Nowak and Allen, 2015). In particular, the author believes that multi-level (including group) selection plays an important role in the evolution of species. This issue is the subject of many articles (see, for example, Wilson and Wilson, 2007). The essence of group selection is that in some cases, the selection within the group must operate differently than on a single individual. According to some scientists, group selection has to play an important role for social organisms, such as social insects. Hierarchy of selection, according to (Wilson, 2007), can be represented as: genes - cell - organism - group.

In particular, papers (Allen et al, 2013, Nowak and Allen, 2015) are devoted to the limitation of inclusive fitness. Inclusive fitness assumes that personal fitness is the sum of additive components caused by individual actions. Authors demonstrated that inclusive fitness is a limited concept, which exists only for a small subset of evolution. According to the authors, this assumption does not hold for the majority of evolutionary processes or scenarios. Currently, however, there is no agreement between different groups of scientists on this problem.

Draghi and Wagner (2010) considered the problem of the evolution of evolvability. According to the author, the problem lies in the fact that evolution (as part of the Darwinian paradigm) cannot be focused on the future. The solution, according to the author, is that the environment is predictable and follows the laws.

Kunin (2011) notes that one of the important concepts of biological evolution - complexity - is badly defined. One of the possible definitions of biosystem complexity is associated with the organizational complexity, i.e., into what organs, tissues, and cells an organism is divided, and how they relate to each other. However, in this case, it is difficult to give any numerical characteristic of such complexity. The complexity of the genome is defined more naturally on the basis of the Shannon entropy. In this case, however, it should be understood that the role of different nucleotides in a sequence is substantially different. This leads to the need to somehow take into account the value of the information.

The author discussed the role of such a mechanism as genetic draft (genetic hitchhiking) in evolution. Genetic hitchhiking, or genetic draft, is the process by which a gene may increase its frequency when it linked to a gene that is positively selected. Proximity of genes on a chromosome may allow them to be dragged along with a selective sweep experienced by an advantageous gene nearby. Genetic hitchhiking can also refer to changes in an allele's frequency due to any form of selection operating upon linked genes, including selection against deleterious mutations. The extent of genetic hitchhiking is closely tied to the rate at which recombination occurs between the mutations.

Kunin (2011) noted that the space of genotypes for even the simplest organisms is extremely large. For example, for prokaryotes with a genome of 1 Mbit, it is possible to have

$$4^{1000000}$$

different sequences. Which part of this amount is actually involved in the evolution? The experimental data (Weinreich et al, 2006, O'Maille et al, 2008) indicate that a plurality of evolutionary trajectories allow only a small part. This means that evolution is largely deterministic instead of stochastic. The author emphasizes that deterministic, in this case, does not mean that evolution has a purpose. However, the question of which mechanisms help limit the space of evolutionary trajectories remains unsolved? Is it limited a priori (with the help of some physical, molecular restrictions) or a posteriori (by means of various forms of selection)?

There are cases when significant evolution occurs in only a few generations or tens of generations (see, for example, cichlids in African lakes, Brawand, D. et al, 2014.). Rodriguez et al (2017) note that consideration of environmental factors is important for understanding the mechanisms of evolution. The problem of evolutionary jumps, in which the changes occur in a relatively short period of time and not gradually, is of particular importance. To solve this problem, the authors (2013, 2015, 2016) proposed to consider the ecosystem in a way similar to quantum mechanics. Based on the large amount of data on ecosystems, the authors proposed the "equation of state" for an ecosystem similar to the equation of state of an ideal gas. According to the authors, these laws should be taken into account in the construction of the modern theory of evolution.

Bacteria demonstrate one of the fastest rates of evolution. The adaptation of bacteria to antibiotics currently represents a major problem in medicine. What would a mechanism for the directed evolution of bacteria look like? Schematically, this can be represented as follows: bacteria, with the help of membrane receptors, determine the state of the environment, including proteins and parts of DNA and RNA, as well as other (and possibly threatening) molecules located in it. As a result of the recognition, the genetic control system produces controlled mutations in the genome.

In recent decades, alternative forms of inheritance were discovered, such as epigenetic processes. The emergence of neo-Lamarckism, the central idea of which is the inheritance of acquired traits, is connected with this direction.

Forms of epigenetic inheritance ('soft') within organisms have been suggested as neo-Lamarckian in nature (Jablonka and Lamb, 2006). In addition to 'hard' or genetic inheritance, involving the duplication of genetic material and its segregation during meiosis, there are other hereditary elements that also pass into the germ cells. These include methylation patterns in DNA and chromatin marks, both of which regulate the activity of genes. These are considered Lamarckian in the sense that they are responsive to environmental stimuli and can differentially affect gene expression. As a result, phenotypic changes occur that can persist for many generations in certain organisms.

Jablonka and Lamb (2006) have argued that there is evidence for Lamarckian epigenetic control systems causing evolutionary changes and called for an extended evolutionary synthesis. According to the authors, the mechanisms underlying epigenetic inheritance can lead to saltational changes that reorganize the epigenome.

How does alternative inheritance change the overall picture of evolution? It is possible, to assume that it is a part of the extended synthesis, but it in no way solves the combinatorial problem. The main issue in relation to neo-Lamarckism is the following: if a priori, the (arbitrary) action of environment on the genome associated with the organism, which is obtained as a result of such changes in the genes. If such a connection is not present, it does not matter that it was the source of mutations. In this case, this effect can be considered part of the Darwinian theory. If such a relationship takes place, i.e., epigenetic inheritance that leads to new benefits to the organism, then it is one of the mechanisms of directed evolution. We emphasize that the question of the specific mechanisms of inheritance (DNA, proteins) in relation to a combinatorial problem is secondary and not fundamental.

It should be noted that in relation to alternative methods of inheritance, the combinatorial problem persists. Moreover, it is exacerbated as the number of possible variants based on genomic methylation became larger. From this point of view, it becomes clear that the combinatorial problem is common to most complex systems, regardless of their carrier of information.

As noted above, the interaction of the organism and the environment must be clearly defined. The term "adaptation" is often used, but this term has many meanings and is poorly defined. In particular, with respect to the epigenetic inheritance, we must determine what adaptation is in this case. Is it different from random (i.e., not directed) changes in the genome and subsequent survival of the organism in the environment? If it differs, then this difference should be explicitly included in the theory of evolution. If not, then such inheritance does not differ fundamentally from Darwinism. If, however, in the organism or in the environment there are mechanisms (programs) of changes in the genome (epigenetics-related or not) that produce the "adaptation", then such mechanisms are part of the general mechanism of partially directed evolution.

Baldwin effect and phenotypic plasticity can be considered mechanisms of speciation.

The Baldwin effect is the theory of a possible evolutionary process that was originally put forward at the end of 19th century. Baldwin proposed a mechanism for the specific selection for general learning ability. Selected offspring would tend to have an increased capacity for learning new skills rather than being confined to genetically coded, relatively fixed abilities. In effect, this theory places emphasis on the fact that the sustained behavior of a species or group can shape the evolution of that species.

The Baldwin effect consists of two steps (Turney et al, 1996). In the first stage, learning during life has a chance to change the phenotype of the individual. If the abilities obtained through learning are useful, the abilities are spread in the population. In the second stage, if the environment is relatively stable, evolution replaces abilities received from learning by congenital abilities (genetic assimilation). Hinton and Nowlan (1987) built the first computer model of the effect. In the literature, there are arguments both for and against the Baldwin effect (Suzuki et al, 2004).

The main problem of the Baldwin effect is that the mechanism by which the abilities are received and become genetically determined remains entirely unclear.

One might speculate that if a priori information about these abilities is absent, then there is only one way by which it is possible to achieve such features during the process of evolution: the exhaustive search. The problem of the characteristic time of the enumeration remains unresolved.

The Baldwin effect is closely related to the concept of "phenotypic plasticity."

Phenotypic plasticity is an important part of modern evolutionary theory. For example, Pigliucci (2007) suggests: "Today we simply can no longer talk about basic concepts like, for instance, heritability, without acknowledging its dependence on the sort of genotype–environment interactions that are best summarized by adopting a reaction norm perspective."

There are two prior opposing theories linking behavior and genotypes. In one theory, conditionally called "behaviorism," the important role of genes was denied. However, there was another trend ("genetic determinism") that stated that genes completely determine behavior.

At present, these extreme views have not been confirmed. The study of the individual development of an organism shows that in different environments, the same genotype may be expressed differently. The conclusion is that the phenotype and behavior of an animal depend not only on genes but also on the environment (see, for example, Agrawal, 2001, Whitam, Agrawal, 2009).

In the opinion of Wagner (2011), phenotypic plasticity is connected with innovations.

The problem with the evolutionary interpretation of phenotypic plasticity is that (as is the case with the Baldwin effect) its mechanisms remain unclear. The mechanism of change in the phenotype must be registered somewhere. If this mechanism is innate (i.e., the genes already included various options for phenotypes in response to certain environmental conditions), then this is simply a variant of evolution. If this mechanism is not inherent, how does a phenotype (behavior) form? For example, because complex behaviors require a large amount of information, the problem of storing this information arises. This need for information storage presents a problem for evolution as a whole and also for many intracellular processes (see, for example, Melkikh, 2013).

On the other hand, the molecular basis of speciation is the work of molecular-genetic control systems (MGCS), or natural genetic engineering systems. How we look at the genome has changed considerably in recent decades (see Shapiro, 2013). Initially, it was only considered a repository of information (read-paradigm), but now it is widely accepted that many of the changes in the genome are caused by the cell itself, or more precisely, by the molecular-genetic control system, which, in essence, is a single unit with the genome (read-write-paradigm).

This control system includes numerous operations, such as mobile genetic element movement, alternative splicing (Will, Luhrmann, 2011, Wahl, Luhrmann, 2015) cutting of DNA, transposons, and others.

It must be emphasized that the molecular genetic control system carries out its work, depending on the environment.

According to Shapiro (2013), there are a number of factors activating genomic instability. Such factors include intercellular signaling molecules and toxic substances. For eukaryotes, a significant correlation between the history of the life of the organism and the epigenetic control system was found. Genome changes in response to stress include point mutations, activation of mobile genetic elements, and restructuring of chromosomes.

All the events listed above suggest the presence of molecular recognition mechanisms, although many of the details of these mechanisms, as well as the work of molecular-genetic control system as a whole, are still not clear.

### 4. Mechanisms of speciation and molecular recognition

Molecular recognition of the environment plays a significant role in the evolution of species. This system should allow the organism to define a state of organisms of their own species (including the closest relatives), and many species with which it is in direct contact. For example, mechanisms for such recognition associated with the operation of the immune system have been previously proposed (see Markov, Kulikov, 2006). The authors note that reproductive isolation has played a key role in speciation. According to the prevailing views, the underlying mechanism of speciation is the gradual accumulation of genetic differences in isolated populations (allopatric phase of speciation) occurring due to mutation, selection and genetic drift. This reproductive isolation was originally conceived as an accidental by-product of adaptation to different conditions (ecological speciation) or the simple accumulation of random changes in the gene pool as a result of a long, isolated existence. In the case of the purely sympatric speciation, it is assumed that the isolation is formed under the direct influence of selection (diverging or disruptive), which favors individuals selectively mated with similar ones. In the paper (Markov, Kulikov, 2006), the possibility of a third variant is justified.

According to the authors, isolation may occur as a byproduct of divergence, but not random, and regular and determined. This may occur on the basis of mechanisms of distinguishing between their own and alien molecules.

Some of these mechanisms may act on the immunological principle by comparing the data on the partner (signaling molecules, pheromones and other antigens in a broad sense) with relevant data about itself. Antigens of the main histo-compatibility complex (MHC) can play a significant role in such testing of potential mating partners. Smell is also involved in the recognition of genetic proximity.

Christakis and Fowler (2014) conducted a study that concluded that on average, the DNA of friends is closer than that of random people in the population. This conclusion is in agreement with the proposed hypothesis that organisms (not just humans) can accurately determine the genotype of another organism. It refers not only to organisms of the same species but also to more distant ones. The study covers a fairly short period of time; however, for the specific time of the formation of new species, the determination of the genetic composition of neighbors could be much more accurate.

Recognition of the molecules in the process of reactions inside the cell is essential for its normal functioning. It is believed that many of the chemical reactions (including those related to the transmission of information) are working on a "lock and key" (or "hand-glove") principle (see, for example, Savir and Tlusty, 2007). This principle is that the shape of one molecule corresponds exactly to the shape of the other. Only in the case of such a complete coincidence does a certain reaction (enzymatic) takes place. If there is no coincidence, then the reaction does not take place, with overwhelming probability.

In the absence of molecular recognition, the stable and precise work of cells would be impossible due to the large number of "abnormal" reactions. Currently "molecular recognition" and "molecular docking" are special directions in molecular biology and biochemistry (see, e.g., Zsoldos et al, 2007, Mobley et al, 2007, Kahraman et al, 2007, Wang et al, 2007). These directions are related to each other and represent structure prediction methods; an example is the prediction of the effective interaction between a protein and a ligand. The peculiarity of this area is that the ligand is typically a simpler molecule than the protein. The authors of many studies note that at present, the accurate prediction of protein-ligand interactions (or more broadly - two biologically important and sufficiently complex molecules) is an unsolved problem.

Let us consider the question of the mechanisms of recognition of the immune system of the state of environment. The immune system makes the following: it organizes the flow of information from outside into the organism. In recognition, such information is first extracted, and then the immune system begins to somehow act on the antigen, but received information may continue its way.

Antigens, as part of environmental organisms, naturally carry information about these organisms. If recognition occurred, then the organism has received information about what organisms surround it. This information includes not only the species of the organisms but also about rather subtle effects of their behavior and evolution, which in some way are reflected in these molecules.

Rather similar processes occur in our very organism. Behavior and evolution, being complex processes, are controlled by a large number of degrees of freedom of macromolecules. Consequently, the opposite is true; reading (recognizing) these degrees of freedom, we can predict the behavior (evolution, biochemistry, etc.) of the environment.

All immune responses of organisms (cells) can be divided into two groups: the organism's immune response as a whole, and intracellular immunity. The work of the immune system of multi-cellular organisms is complicated and includes so-called innate immunity and acquired immunity. The most important property of the immune system is the ability of an antibody to selectively bind to an antigen. The mechanism of this selectivity remains largely unclear, as well as the temporal evolution of antibodies. The intracellular immunity is related to RNA interference (see, for example, Castel and Martienssen, 2013, Jaronscyk et al, 2005).

Note that the problem of recognition of antibodies in the immune system is relevant to general biological and medical applications. It is believed that this problem is solved by so-called "housekeeping genes", and their strong variability. However, mathematical models of the recognition process are absent in the literature, and the problem of the number of variants of antibodies and their possible enumeration in terms of complexity is not discussed.

As research has shown, human and animal senses allow them to distinguish much more subtle effects than previously thought. For example, humans can distinguish between 2.3 to 7.5 million colors (Bushdid et al, 2016, Dickerson, 1943) and approximately 340,000 tons of sound. As for the smell, the number of different variants for such signals is estimated at approximately $10^{12}$ (see, also, Meister, 2015). These values (related, of course, to other animals) are difficult to justify as adaptations for survival.

Moreover, certain species (e.g., *Drosophila*) even distinguish isotopes (e.g., hydrogen) (Franco et al, 2010). This can significantly increase the possibilities of the olfactory system.

These estimates provide a lower bound for the capacity of the information channel of external information processing. To give an upper bound is much more difficult because recognition at the molecular level is much more difficult to track in the experiment. However, autophagy allows us to make such estimates. Autophagy (see, for example, Mizushima et al, 1998, Ichimura et al, 2000) is a system for combating cellular "trash". Autophagosomes can detect incorrectly folded proteins and other molecules. This leads to very different estimates of possible degrees of freedom that can be recognized and for which appropriate controllable actions can be performed. Generally speaking, a protein in a conformation different from the native one represents another molecule. Of course, this refers also to other biologically important molecules (RNA, DNA).

This issue was examined in the work (Melkikh, Seleznev, 2012). Because different proteins interact with each other, it is unclear how to address the avalanche of errors. If the newly formed complex will lead to some new reactions, do they in turn lead to other new reactions? Protein networks form, where proteins can fall not into their native configurations, but in some long-lived local minima, where they are notoriously inefficient.

How do autophagosomes distinguish the wrong protein complexes and incorrectly folded proteins? For such a distinction, autophagosomes must have receptors for specific proteins. In principle, such receptors might exist for all proteins, but there are thousands of different proteins in a cell. Even in this case, there is a problem because the number of molecules of each receptor would be only a few if not one. What would be the speed of information processing in such a system?

However, the main problem is not this, but how to recognize the different conformations of protein. The number of such possible conformations (taking into account the total number of proteins) is exponentially large, so space for all receptors is simply not enough. To only recognize the right proteins, it is necessary to have a mechanism of such recognition. How should a receptor be arranged that only recognizes one type of protein and eliminates any "wrong" proteins, but does not react with the "correct" form of all other types of proteins (thousands of them)? Most likely, such a receptor will have to present very stringent requirements in terms of complexity. That is, it will have to have a very large number of controlled degrees of freedom.

It can be assumed that the immune system and other sensory systems of the organism an important part of the work carry out in the framework of directed evolution. That is, directed evolution requires, as much as possible, accurate knowledge of the environment down to the molecular level. This is because the organism can determine which neighboring niches are occupied and which are free. Information capacity of all senses (including immunity) can completely solve this problem. That is, the senses determine exactly what organisms are found in the environment; on this basis, such (directed) changes in the genome occur, in accordance with which organs (such as wings) gradually begin to arise, allowing the organism to survive in a specific environment. However, the basic problem – the way of processing of this information – has not yet been solved. Solution of this problem would play the crucial role in creation of novel evolution theory.

## 5. Quantum effects of the interaction of biologically important molecules and mechanisms of speciation

We first discuss the problems of the work of the molecular-genetic control system. Next, in this context we discuss mechanisms of speciation.

The question of the mechanism of the molecular-genetic control system, as well as the biochemical reactions in the wider sense, was discussed earlier in (Melkikh, 2014, 2015), mainly in relation to the problem of protein folding. However, it should be noted that folding of DNA and RNA plays an equally important role in the work of the molecular-genetic control system. For example, it is known that DNA is tightly packed into the cell nucleus. In this case, in addition to the package as such, there is a problem of access to the different portions of the DNA for their regulation.

In a cell, DNA is folded into nucleoprotein structures. When forming the mitotic chromosome, the DNA of eukaryotic cells is folded several thousand times with great accuracy (see Gatti, 1983). However, despite the great progress in the study of DNA folding, the mechanism of such precise DNA folding remains unclear.

In eukaryotes, DNA is condensed into chromatin. Between cell divisions, chromatin is optimized for accessing active genes. However, it remains unclear how such selective access occurs. During division, the chromatin is folded in classical chromosomes, where DNA is structured at a higher level.

We show that the problem of DNA folding, as well as its function as a repository of information, is not only nontrivial, but is also even more contradictory and paradoxical than protein folding.

Consider compact DNA folding in chromosomes. As we know, DNA in the chromosomes represents a condensed medium of high density (see, for example, Teif, Bohinc, 2011). Such a medium can be obtained only in the case of a high orderliness of the polymer chains. Let's estimate the number of degrees of freedom of the DNA folded in chromosomes. Let the double helix have a length of 3 billion base pairs of nucleotides. Even if we take as the domain the persistence length of the polymer (equal to approximately 50 nm), the total number of folding variants of such a polymer would exponentially large. Indeed, if we accept that as a result of bending, a DNA chain on the characteristic persistence length can take at least two different states, then we obtain for the total number of possible spatial states of DNA:

$$2^{L/L_{pers}}.$$

This is a lower estimation, but with enzyme action this length can be made much smaller. However, even this estimation gives a number approximately equal to:

$$2^{2\times 10^7}.$$

This number of variants is so large that it is impossible to enumerate them during the lifetime of the universe with parallel operation of all living beings who ever lived on Earth. This means that the DNA during folding has become entangled in any of the exponentially large numbers of amorphous states.

In relation to the condensed DNA, this means that each nucleotide is surrounded by approximately six other nucleotides, but known potentials of interaction between the atoms have a characteristic length of the order of atomic sizes. In this case, the misfolded structure will correspond to the potential well in which the system will stay long enough. On the other side, there are exponentially more such potential wells (this shown, for example, for the spin glasses). However, potentials for other molecules (e.g., proteins), which would initially prevent creation of the wrong spatial structure of DNA, are not known.

If for relatively short proteins, the mechanism of simple enumeration of variants could work during folding process, then for DNA it is impossible because of its significantly larger size.

It is believed that the control of the folding of DNA by proteins solves the problem of folding (e.g., histones promote folding of DNA into nucleosomes); however, proteins are also molecules for which the same paradox occurs, which means that instead of having to control the folding of DNA in regular structures, such proteins could entangle DNA because the number of entangled states is much larger than the number of correctly folded states.

The problem of DNA and RNA folding from the physical or geometrical point of view has been little studied in the literature. However, it is obvious that most of the models and their contradictions discussed above are fully applicable to these structures. That is, the DNA and proteins, performing any operation on it, represent one large macromolecule, for which folding is contradictory.

There are many papers devoted to the problem of protein folding (Bryngelson and Wolynes, 1987, Onuchic, Volunes, 1997, Volunes, 2004, Volunes, 2015, Grosberg, Khokhlov, 2010, Finkelstein, 2013, Martinez, 2014,

Ben-Naim, 2013, Berger, 1998, Bern and Bayes, 2011, Crescenzi et al, 1998, Shaw et al, 2014), but a general solution of the problem has not been obtained.

As it was noted above molecular docking and molecular recognition is a separate area in which the configuration is calculated when the protein and ligand interact. It uses some simplifying assumptions, smoothing the energy landscape. We emphasize that only the geometric approach (Berger, 1998, Crescenzi et al, 1998, Shaw et al, 2014) can be considered the one from first principles because only in this approach a specific topology of biologically important molecules is taken into account. In many other cases, it is either ignored or simply a smoothed energy landscape. However, the task becomes significantly different and no longer corresponds to what happens in nature.

Since the problem of folding and reactions is NP-hard, i.e., requires an exponentially large number of steps, this leads to a contradiction with characteristic speed and accuracy of these processes. As shown in (Melkikh, 2014, 2015), the main processes in the cells will not happen, as the proteins will not have time to take their native conformations and biologically important molecules in the interaction will become entangled, forming inefficient complexes, and making genome control impossible.

All the above leads to the need of the application of quantum mechanics for modeling the interactions of biologically important molecules in general and the molecular-genetic control system in particular.

Let's consider the problem of origin of biological complexity and mechanisms of speciation.

Considering the allele variants, not all of the possible nucleotide variants significantly change the problem. The number of variants of nucleotides is, of course, much larger. However, this consideration is only a posteriori (that is, comes from what realized), not a priori, which must come from what is permitted by the laws of nature.

Such well-known methods to accelerate evolution compared to the exhaustive search of variants, as cumulative selection and block coding require a priori information for their implementation.

Under cumulative selection, the information sequence is iterated at random until a nucleotide reaches the "good" (goal) state. This search method was proposed by Dawkins (1986) using the example phrase "Methinks it is like a weasel".

To show that this method requires *a priori* information, we consider in more detail how "good" nucleotides "fix"; without this process, the sequence can be destroyed by random processes. If there is a special mechanism of protecting "good" nucleotides from mutations, there must be a mechanism of recognition. In turn, identification (according to the recognition theory) requires standards for comparison. Such standards must exist before the recognition process and represent no more than *a priori* information about these nucleotides encoded in some structures. It is easy to show that the amount of this information will be approximately equal to 2N bits (as in the above proposed algorithm, partially directed evolution, where $K \approx N$).

Assuming that the external environment somehow directly or indirectly affects the nucleotide sequences, then it is necessary to determine the nature of this external environment. It is something active, or some machine? If yes, then we must determine the sequence of actions of such a machine; if not, then there is no reason to believe that it would select something with blocks. If the external environment has not a "target" to create organisms adapted to anything, then its role will only be organism survival.

An alternative method of nucleotide fixation might consist in the arrangement of the space of adaptation such that there always exists a path from one niche to another. That is, selection favors only those organisms that are on this pathway. In other words, those organisms in which some nucleotides have reached the target set should certainly survive; however, unless the further mutation of these nucleotides is prevented, selection will not be able to protect them.

Let us show that this is so (here we follow Melkikh (2014)). Indeed, to preserve any set of nucleotides unchanged (while others are subject to change) in the absence of an internal mechanism, but only by selection, we must have a population of organisms comparable to the total number of states of these nucleotides. If there is an exponentially large number of nucleotides (and there will be more and more of them in proportion as cumulative evolution continues), then this number is also exponentially large.

We can conduct a physical analogy with Brownian motion: if every particle that has moved beyond the specified area is destroyed, then the time of existence of particles within it will be small. It can be large only when an exponentially large number of particles exists. If this area is bounded by a potential barrier, this situation corresponds precisely to the internal mechanism of recognition, which is discussed above.

In addition, cumulative selection requires a special *a priori* arrangement of the environment (adaptation space). In this case, the question arises: how could such an arrangement appear?

Note that the speciation mechanisms discussed above are, in fact, not microscopic mechanisms, but macroscopic schemes requiring microscopic grounding.

However, in this case, how is the choice made from the exponentially large number of variants? It is believed that during evolution, organisms do not enumerate all possible variants, but are restricted by repetitive use of the same (or related) sequences. For example (Putnam et al. 2007), mammalian genomes are composed of a considerable part of almost similar genes. This process of using old sequences for new purposes is often called molecular exaptation.

Let us ask the question of in what case is such exaptation possible? This issue is discussed in the work (Melkikh, 2015), in which it was concluded that such a process must inevitably be accompanied by the presence of a priori information about how good a given sequence will be. The presence of such information radically changes the mechanism of evolution; from the non-directional (Darwinian), it becomes directional. One would assume that exact solutions are not realized in nature, but always only approximate ones, i.e., many nucleotide sequences could have some meaning, or encode any organism. Indeed, within each species there is some genetic diversity within the population of several different organisms. However, such diversity (some degeneracy of solution) essentially does not alter the conclusion that in the case of non-directional evolution, the problem of enumeration of variants remains NP-complex. The paper (Melkikh, 2008) calculated the probability of the formation of species in the frame of the model of undirected evolution, which explicitly takes into account intraspecific and interspecific differences, which are on average 1/1000 and 1/100. It was shown that the probability of such species origin on such a mechanism remains exponentially small, and NP-complexity weakly depends on the accuracy of these quantities.

Evolution of species can be described on the basis of known system of equations

$$\frac{dx_j}{dt} = x_j \left( \sum_{i=1}^{n} a_{ji} x_i - \varphi(t) \right), \qquad (1)$$

where

$$\varphi(t) = \sum_{j=1}^{n}\sum_{i=1}^{n} a_{ji} x_i x_j, \quad \sum_{i=1}^{n} x_i = 1.$$

The values of $x_i$ are the frequencies of alleles in a population and $\varphi(t)$ is the fitness.

The matrix $a_{ij}$ of mutational flows is considered symmetrical. Fisher (1930) showed that the change in the average fitness is non-negative (see, also, Schuster, 2009):

$$\frac{d}{dt}\varphi(t) = \sum_{j=1}^{n}\sum_{i=1}^{n} a_{ji}\left(x_i \frac{dx_j}{dt} + x_j \frac{dx_i}{dt}\right) = 2\sum_{j=1}^{n}\sum_{i=1}^{n} a_{ji} x_i \frac{dx_j}{dt} =$$

$$= 2\sum_{j=1}^{n}\sum_{i=1}^{n} a_{ji} x_i \left(\sum_{i=1}^{n} a_{jk} x_j x_k - x_j \sum_{k}\sum_{l} a_{kl} x_k x_l\right) =$$

$$= 2\sum_{j}^{n} x_j \sum_{i}^{n} a_{ji} x_i \sum_{k}^{n} a_{jk} x_k - 2\sum_{j}^{n} x_j \sum_{i}^{n} a_{ji} x_i \sum_{k}^{n} x_k \sum_{l}^{n} a_{kl} x_l = \quad (2)$$

$$2\left(\left\langle\langle a\rangle^2\right\rangle - \langle\langle a\rangle\rangle^2\right)$$

The result (Fisher's theorem for evolution) suggests that the population, on average, is moving in the direction of the local peak of fitness.

We make a few remarks on this theorem.

First, as noted above, it is not obvious that in the framework of evolution simulation we must consider the alleles of genes. It is necessary to consider all the genes changes that are not prohibited by the laws of nature, that is, arbitrary permutations, or insertion and removal of nucleotides. This consideration is in the spirit of Darwinism, as while descendants are not created, in the framework of undirected evolution it may not be aware how good the nucleotide sequence is. Consideration of alleles alone significantly reduces the number of possible variants, but is based on the implicit assumption that other variants of the genome are prohibited a priori. This assumption may not be justified as part of undirected evolution.

Second, genes are linked with each other. In this case, according to (Schuster, 2009), the optimization principle is inconclusive. However, gene connectivity should be more precisely defined. The paper (Melkikh, 2014a) examined the field of mutations and field of fitness, which can be given a geometric interpretation. On the plane can be identified two different vectors: gradient of fitness and mutation vector. Without additional assumptions, it does not follow that they are directed to one side because they depend on very different parameters.

If there is no correlation between these vectors, i.e., the average projection of the vector of mutations on the gradient of fitness is zero, this evolution can be called truly random (non-directional a priori). If such a correlation exists, than evolution is partially directed. At full correlation (i.e., when the mutation a priori is directed towards increasing adaptability), evolution is entirely directed. Note that in the case of non-directional evolution, mutations can be arranged arbitrarily complex, i.e., it does not require that the different nucleotide mutations were equiprobable and independent. It is sufficient that this complex process of mutations was *a priori* not focused on the field of fitness.

Thus, within the Darwinian paradigm of undirected evolution (including extended synthesis), the theorem remains true, despite the connectivity of genes (nucleotides).

In the frame of partially directed evolution movement in phase space can be described by Fokker-Planck equation (Melkikh and Khrennikov, 2016):

$$\frac{\partial f}{\partial t} = \vec{v}\vec{\nabla}f + D\Delta f ,$$

where *v* is a velocity vector of directed motion due to a priori information, *f* is the probability density, and *D* is the diffusion coefficient.

Such a directional movement can solve a known problem of crossing "ravines" of fitness. Indeed, in the framework of undirected evolution, organisms that reached the nearest peak of fitness find themselves in a "trap", and the achievement of a neighboring peak of fitness (even if it is much higher than the maximum on which they are located) is highly unlikely. Within the framework of the theory of undirected evolution, it is not possible to offer a consistent mechanism of crossing ravines.

In the framework of directed evolution movement towards the highest set of fitness is just a part of this theory, in which the movement down the slope of adaptation is not prohibited and is not unlikely (Figure 1).

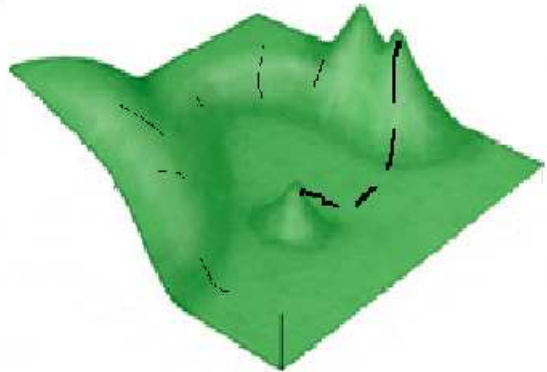

Fig.1.

With regard to the formation of species, it is appropriate to also consider the question of the origin and existence of the genetic alphabet. Why is there a single genetic alphabet? Why there are alleles of the genes in the population? From the standpoint of the Darwinian theory, it is considered an axiom that requires no proof. We show, however, that the existence of alleles itself leads to a contradiction in the frame of undirected evolution.

Indeed, let the genetic alphabet has arisen for some reason. That is, all creatures have the same rules of reading genes, determining the beginning and end of genes, etc. How will such a system evolve? Let one character in the alphabet be changed. That is, any nucleotide now means something new (it can be nothing). However, in terms of undirected evolution, it does not mean anything. This descendant will survive, or not; the mechanism of gene changes does not matter. This means that the genetic alphabet itself, as well as conformity between nucleotides and amino acids, is neutral.

Then, an alphabet should change over time, as all its rules and commands can be executed in a very large number of ways. In the modern theory of evolution, it is assumed that since both the genetic alphabet and the genetic code are very important, therefore they are also very conservative (they change little in the course of evolution). However, it should be specified for whom (to what a structure) these properties are important. They may not be important for the organism itself, but only for the descendants; however, the descendants' survival is not directly connected with the rules of the genetic language. The terms of undirected evolution first require a descendant to be made, and then how it is adapted to live in such an environment will be determined.

To ensure the sustainable existence of the genetic alphabet with all its rules, the existence of a special genetic control system is necessary. This system determines that this gene is dominant, but not recessive, and that the definite sequence of nucleotides serves as a punctuation mark between genes. The existence of this control system cannot be justified within undirected evolution, since it does not provide advantage to the organism by itself. If we assume that the system arises and functions accidentally, then again there is the problem of the enumeration of an exponentially large number of variants, which in the frame of undirected evolution cannot be solved.

From the viewpoint of partially directed evolution, the existence of a control system of genes and their evolution is natural. This process can be compared with the growth of trees: when we planted a birch seed, we knew in advance that a birch will grow, and not pine, for example. What exactly will grow depends on external conditions (humidity, soil, light conditions), but there is no doubt that this is a controlled process that is largely controlled by genes. The term "growth program" can be used in this sense. This program provides various development variants, which are run by a certain state of the environment, determined by receptors.

Similar mechanisms should be implemented in the evolution of species. To confirm or disprove the existence of such a mechanism, special experiments required. The most promising may be considered experiments with rapidly evolving species. It is not sufficient to determine the state of genes within a certain period of time, but we must follow evolution "on-line", that is, the process of change in the genes at characteristic molecular times. This is difficult task, but possible in perspective. For a more detailed discussion of evolutionary experiments, see (Melkikh and Khrennikov, 2015).

Thus, from the point of view of partially directed evolution are the following *requirements* for the mechanisms of speciation:

First, there must be a system of molecular recognition of the environment.

Second, the information obtained can be used for the start-up of existing alleles (operational program of evolution) and to create the operational program itself. That is, in a directed way, alleles that may be needed in the near future will be produced during reproduction.

Third, the crossing itself is directional. Wherein selected a partner, who presumably has genes that are most appropriate to the environment.

Fourth, at crossing, spare genes will be produced, which can be useful for population in the near future.

Fifth, the life span is adjusted so that if necessary, rapid changes in the population (when the environment has changed significantly over the generations) free up space for more adapted descendants (see, also, Melkikh and Khrennikov, 2016).

When part of the population became isolated from the rest of the population (for whatever reason), then spare programs can survive (i.e., those that are not the most optimal). In this case, a set of other species arises. In the limit, when a small part of the population became isolated at the edge of an areal (i.e., under special conditions), then their a priori program practically did not have to compete with each other, and one of them survived.

In the absence of spatial isolation, environmental recognition by many organisms gradually happens. Operational programs accumulate, until one day the program that provides an effective view for this niche will run. During this transition, a new species ceases to interbreed with the old one. This process is also controllable and predetermined.

Baldwin effect can also be explained by the theory of partially directed evolution. It is important to note that the behavior, including learning, is also due to a priori information (see also Melkikh, 2011, 2014c), which means that the behavior, as well as the evolution, is a priori-directed (for more on the similarity of behavior and evolution, see (Melkikh, Khrennikov, 2015)). In this case, it can be concluded that it may be advantageous under certain conditions to implement inherent behavior as a program that is recorded in some structures (not necessarily in the genes) and in other cases as genes that encode the behavior itself. An advantage of this or any other type of recording of information can depend on the uncertainty (stability) of the environment.

That is, in some cases, from the multitude of bits that form the program, it is necessary to reserve a set with much less power that will be used in the future for the selection of different behaviors. Naturally, the choice of this set is based on a priori information about the evolutionary landscape. This hierarchy in the programs of behavior is not limited only to living organisms but is also characteristic of many technical systems.

Note that the implementation of the Baldwin effect in models of artificial life clearly shows that this effect can be realized only with a priori information. For example, according to Red'ko (Red'ko et al, 2005), genes in a population of artificial organisms set neural network parameters. Neural networks can be optimized by evolution as well as by learning. It is important that the properties of neural networks are genetically determined, i.e., explicitly present in the genome. Thus, in this case, the Baldwin effect is nothing but a type of evolution.

Note also that the Baldwin effect is used in genetic algorithms and hybrid models. In these cases, the effect is a symbiosis of neural networks and genetic algorithms. However, genetic algorithms and neural networks involve a priori information, and without a priori information, the genetic algorithms and neural networks cannot work.

Thus, the concept of "phenotypic plasticity," "plasticity of behavior" and the "Baldwin effect" can be interpreted based on the theory of partially directed evolution as follows:

- there exist behavioral programs and organism structures, which are recorded either in the genes or in other structures;

- a part of these programs is stored in latent form and is not working;

- with changes in the environment, certain programs of the behavior (or the phenotype change) run, but the genetic part of programs may not change;

- when the environmental changes become stable, it is advantageous to change the genetic composition, but again, this is done in accordance with a priori information recorded in some structures.

As noted above, quantum effects should play an important role in evolution. In accordance with the general ideology of control theory (see Appendix), write the fundamental system of equations for the fields and particles in the following form:

$$\left\{\gamma_\mu\left(\frac{\partial}{\partial x_\mu} - ieA_\mu(x)\right) + m\right\}\psi(x) = u\psi(x)$$

$$\left\{\gamma_\mu^T\left(\frac{\partial}{\partial x_\mu} + ieA_\mu(x)\right) - m\right\}\bar{\psi}(x) = u^*\bar{\psi}(x)$$

$$\Box A_\mu(x) = -j_\mu(x) + vA_\mu(x)$$

where $\psi(x)$ is the wave function (operator) and $j_\mu(x)$ is the 4-density of electron current, which is equal to

$$j_\mu(x) = \frac{ie}{2}\left(\bar{\psi}(x)\gamma_\mu\psi(x) - \bar{\psi}^c(x)\gamma_\mu\psi^c(x)\right)$$

The subscript "c" indicates the change of the charge sign. $A_\mu$ is the potential of the electromagnetic field, $\gamma_\mu$ are the Dirac matrices, $e$ is electron charge, and $\Box$ is

$$\Box \equiv \frac{\partial^2}{\partial t^2} - \frac{\partial^2}{\partial x^2} - \frac{\partial^2}{\partial y^2} - \frac{\partial^2}{\partial z^2}.$$

Here $u$ and $v$ are controls – they are components of united vector of control:

$$\begin{pmatrix} u \\ v \end{pmatrix}$$

In the absence of the control, the system (4-6) is used in quantum field theory (Akhiezer and Berestetskii, 1965) within the framework of the second quantization.

The cost function in general will depend on basic variables and controls:

$$I = I(\psi, u, v).$$

A specific type of cost function should be determined on the basis of specially designed experiments.

Note that in light of the above, the interaction of biologically important molecules, a hallmark of the interaction between biologically important molecules must be their non-locality. For the simulation of systems with such properties, classic (Li, 1992) and quantum cellular automata (see, for example, Elze, 2015) have been used.

A related issue is the problem of the growth of quasicrystals in which the interaction between groups of atoms is essentially non-local to obtain the right quasicrystal (see, Bindi et al, 2009, Steinhardt, 2008, Marcia, 2006).

Another example of non-local interactions is neural networks (live or artificial), in which each neuron can communicate not only with its neighbors but also with arbitrarily remote neurons. There are also models of quantum neural networks (see, for example, Gupta, 2001).

The paper (Melkikh, 2014b) proposed a model of quantum motion control of biologically important molecules. One of the main provisions of the model is to allocate the part of the Hamiltonian associated with many-particle interactions with sufficiently distant particles. It was hypothesized that a field φ is responsible for this interaction, which itself can depend on the time and parameters of the system. We can write the equations of the model in the following form:

$$i\hbar \frac{\partial \psi}{\partial t} = \hat{H}\psi + \varphi \psi \tag{8}$$

$$\frac{\partial \varphi}{\partial t} = g(\varphi, \psi). \tag{9}$$

The first equation is the Schrödinger equation for a particle which, besides the usual Hamiltonian, also contains the potential, which corresponds to the collective interaction of particles.

The second equation represents the dynamics of the many-particle potential. This particular potential organizes collective effects so that the folding of proteins and other reactions between biologically important molecules occurred in the funnel-like landscape.

According to (Melkikh, 2014b), a zero solution leads to the standard Schrödinger equation. A nonzero solution leads to nontrivial quantum effects in relation to biologically important molecules.

The role of the function g ($\varphi$, $\psi$) in the right part of equation (9) is that under certain conditions, it allows for the ban on most degrees of freedom, except for only a small set. These are exactly those degrees of freedom that enable rapid protein folding, effectively implementing the reaction of "lock and key"-type and other cellular processes discussed above.

It was noted in (Melkikh and Khrennikov, 2015) that quantum effects are not necessarily associated with wave behavior. For example, magnetism is essentially a quantum phenomenon; however, it shows no wave properties. This means that the presence of quantum effects in the interaction of biologically important molecules does not necessarily need to be associated with their wave behavior. In general, these molecule masses are too large for their wave behavior.

The molecular recognition problem discussed above may be considered as well, and in more general terms. It can be shown that only innate programs can be run as a result of recognition. This issue was previously discussed in relation to the problem of knowledge acquisition in the broader context (Melkikh, 2014). Indeed, as we know from the theory of pattern recognition, the recognition process itself is possible only if there is an etalon (reference pattern) for comparison. This is also implemented in neural networks with the teacher. That is, to recognize the object - to attribute it to the certain class - it is possible only when a priori properties of this class are defined.

Therefore, recognition by the receptor of a molecule is only possible if its type is known in advance. Otherwise, a signal that produces a receptor in output will not belong to any a priori class. Consequently, such a signal will not be useful.

Thus, the surrounding environment recognition processes by the immune system (as well as other sensory systems) are in any case congenital, i.e., they are part of program of evolution. An algorithm of evolution comprising a quantum step can be represented as follows:

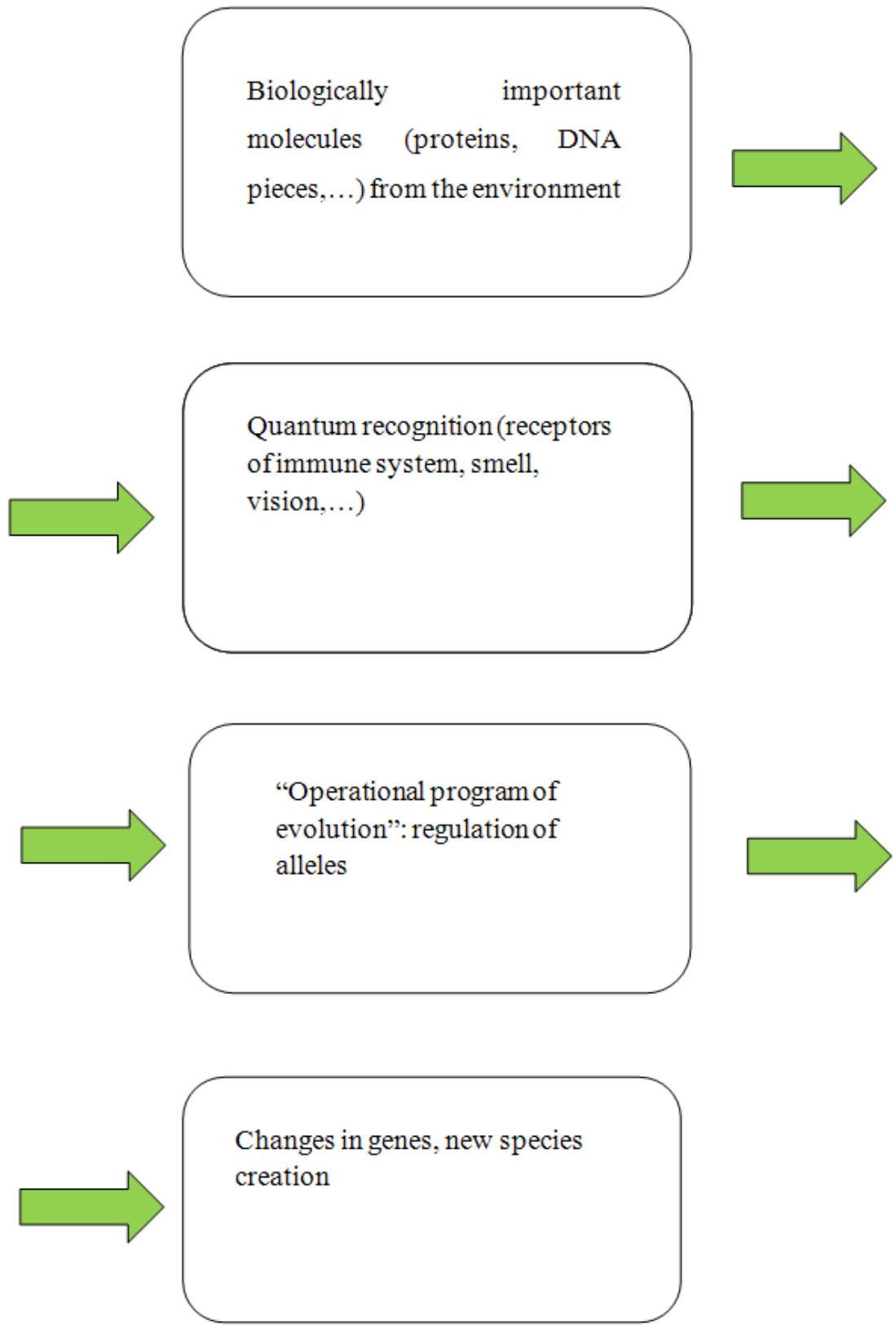

Fig.2.

Thus, the genome eventually changes (including epigenetic changes). Gradually, new species and subspecies arise.

For example, only 1/6 of the yeast genome comprises genes essential for life (Hillenmeyer et al, 2008). Others represent different alleles of the same gene, which may be useful in the changing (non-ideal) circumstances. The same can be said about the genomes of all living beings. From a Darwinian point of view, it is impossible to justify why the organism stores some genetic sequences "in reserve". Organisms will not profit by it, and this maintenance requires material and energy.

Consider some possible experiments that could clarify the actual mechanisms of speciation (see, also, Melkikh and Khrennikov, 2015):

- Experiments on DNA folding and allocation of chromosomes in the cell. A systematic study of the interaction forces between the chains of DNA (RNA, proteins) of different composition could shed light on the nature of the interaction between the biologically important molecules. For this, high-speed ($10^{-15}$ s) laser spectroscopy to measure the intermediate states of molecules can be used.

- Experiments with rapidly evolving systems. As was noted above, one of the fastest evolving systems is bacteria. What mechanism of mutation occurs in such cases? It is necessary to track the entire logical chain in this evolution, starting with the reception of the environment, followed by the processes of intracellular regulation of the genome to mutations that lead to the emergence of new genes. Eventually, we must consider such a chain of processes, and any a priori information stored in some intracellular structures will be found.

Experiments on epigenetic effects can also play important roles. It is necessary to determine the mechanisms by which particular methyl groups are attached to a specific location in the genome. This can be done on the basis of NMR techniques and using different hydrogen and carbon isotopes.

## 6. Quantum bioinformatics and the problem of epigenetic plasticity

In a series of works of Asano et al. (2010, 2011a,b, 2012a,b, 2013. 2014, 2015a,b) the mathematical formalism of quantum theory was applied to model adaptive behavior of biological systems, see also (Melkikh and Khrennikov, 2015) for review. This approach is known as *quantum bioinformatics.* In this section we want to present briefly the basics of this novel approach by emphasizing its difference from quantum biophysics and the classical bioinformatics, see (Asano et al., 2015b) for details and discussion. Then, in section 6.6, we apply this approach to modeling of the epigenetic plasticity.

### 6.1. Theory of open quantum systems and biological adaptivity

We emphasize that biological systems are fundamentally open systems, i.e., they cannot survive as isolated systems. Therefore, to model their behavior, it is natural to apply theory *of open quantum systems* describing dynamics of a system *S* interacting with a bath (reservoir) *E*.[1] The latter has a huge number of degrees of freedom. It is difficult (if possible at all) to describe explicitly the dynamics of the compound system *S+E.* Therefore quantum theory of open systems explores approximations of the complete state dynamics. The most widely used is the

---

[1] See, e.g., Asano et al. (2015) for biologist friendly presentation of theory of open quantum systems and more general theory of adaptive quantum systems. A more advanced presentation, but also with biological flavor, can be found in (Ohya and Volovich, 2011). And, finally, the rigorous mathematical formalism is presented in the book (Ingarden, Kossakowski and Ohya, 1997).

Markov approximation in the form of *Gorini-Kossakowski-Sudarshan-Lindblad* (GKSL) equation for the dynamics of the *S*-state.

In works of Asano et al. (2010, 2011a,b, 2012a,b, 2013. 2014, 2015a,b), it was pointed out that theory of open quantum systems and more general theory of quantum adaptive systems (Asano et al. 2015a) can be used to model the state dynamics of not only quantum systems, but even biological systems. In such modeling, *S* is an arbitrary biological system, from genomes, proteins, cells to animals and ecosystems, and *E* is the surrounding environment. Thus the quantum master equation in the approximate form of the GKSL equation describes the evolution of the state *ρ(t)* of *S*. Here "state" is treated as the *information state* of *S*. Information encoded in *ρ(t)* is not reduced to the information about the physical parameters. It includes also information about the biological degrees of freedom of the system *S* and the environment *E*. The impact of the latter is encoded in the coefficients of the GKSL equation, in the so called *Lindblad operator* of this equation, denoted by *L*.

We remark that the generator *G* of this equation consists of two terms, Hamiltonian *H* describing the intrinsic dynamics of the state of *S* and the Lindblad term *L*. The latter generates adaptation to the environment. Thus *G=H + L*.

In contrast to the Schrödinger equation, the GKSL equation does not preserve the purity of a dynamically evolving state. A pure initial state *ϕ* can be immediately transformed into a complex mixture of a few pure states (Ingarden, Kossakowski and Ohya, 1997). Mathematically such a mixture is given by a *density operator ρ*. Thus this equation describes the dynamics of the density operator, *ρ (t)*.

For a "natural" Hamiltonian *H* and Lindblad operator *L,* the state *ρ (t)* stabilizes to some *steady state ϑ*. This state represents a stable configuration for the biological system *S*. (The latter can be a genome, epigenome, cell, animal, human being, ecosystem, social system.) In the mathematical model (Ingarden, Kossakowski and Ohya, 1997), stabilization of $\rho(t)$ to $\vartheta$ is takes place for $t$ approaching infinity. Of course, in reality this stabilization takes finite time: after some time interval fluctuations of system's state $\rho(t)$ become relatively small, they can be ignored and $\rho(t)$ can be treated as approximately equal to $\vartheta$.

Changes in the environment will modify the Lindblad operator $L´$ and, hence, the dynamics of $\rho(t)$. The steady state $\vartheta$, the output of interaction of $S$ with the previous state of environment, loses its stability (in the new environment $E´$). Starting with $\vartheta$, the new dynamical describes the evolution of system's state. The latter will sooner or later stabilize to the new steady state $\vartheta´$ and so on. Since the generator $G$ of the adaptive evolution described by the GKSL equation has the form $G=H + L$, the dynamics can be modified not only due to changes in the environment, $E \rightarrow E´$, but as the result of changes in system's Hamiltonian, $H \rightarrow H´$. The latter represents the internal "information energy" of $S$, the genuine potential of a biosystem to change its state.

## 6.2. Principle of complementarity in quantum physics and biology

One of the main distinguishing features of the quantum evolution is its *probabilistic character*. The symbolic expression "system's state" represents statistical features of a population of systems. A state $\rho$ (even a pure state $\phi$) determines only the probabilities for expressions of some physical and biological features by representatives of the population. This is the essence of the so-called *statistical interpretation of quantum mechanics.*

However, by itself the probabilistic nature of the state dynamics is not the distinguishing feature of the quantum theory. By considering, instead of the quantum master equation (e.g., in its Markovian form - the GKSL equation), the classical master equation, we shall also obtain probabilistic dynamics. One of the main distinguishing characteristics of the quantum theory is that here the state of a system encodes its *incompatible features*. Such features cannot be exhibited by *S* simultaneously. But potentiality of their realization is present in system's state.

As an illustration of such incompatibility, we point to gene expression. Incompatible expressions coexist in genome; we can concretely point *to lactose-glucose metabolism*, see Asano et al. (2012b, 2013, 2015a), Basieva et al. (2012) for its quantum-like treatment.

We can also present a plenty of examples of incompatible mental expressions of human behavior. For instance, we can point to *the disjunction effect* playing the important role in cognitive psychology and illustrating irrationality of human behavior. This effect demonstrates that behavior of humans in context of uncertainty is incompatible with behavior in context of resolution of uncertainty; see (Khrennikov, 1999, 2004a, 2004b, 2010, 2015, Conte et al., 2004, 2007, Busemeyer et al., 2006, 2011, 2014, Busemeyer and Bruza, 2012, Haven and Khrennikov, 2012, Pothos and Busemeyer, 2013) for quantum modeling of this effect.[2] Following these authors, let us consider the following experiment for psychological behavior.

---

[2] The second author of this paper approached this problem by starting with probabilistic analysis of quantum foundations. The main output of this analysis was that quantum probabilistic behavior has no rigid coupling with some "mystical features" of micro-systems. Supported by this conclusion, Khrennikov started to look for applications of quantum probability theory outside of physics, especially to model cognitive and psychological behavior. In this way there was established the fruitful cooperation with the group of experimenters working in cognitive science (under the leadership of E. Conte). E. Conte proposed to test quantum-like features of statistical data collected in experiments on recognition of ambiguous figures (Conte et al., 2004). There was explored the experimental design proposed in (Khrennikov, 2004a,b). Honestly speaking publications (Conte et al., 2004, 2007) did not attract so much attention. The revolutionary step was done by the professor in cognitive psychology J. Busemeyer who approached the same theory from another side. He had been working for long time with disjunction effect and he was interested to find novel mathematical machinery to handle this effect in the proper way. And he started to appeal to quantum probability. As the result of his advertising of quantum probability in the cognitive psychology community, the quantum(-like) models started to diffuse (still very slowly) into this community.

A group of people participating in the experiment should play a game in which they can either earn 200 USD or lose 100 USD, the probabilities of such outputs are equal, *p(200)=p(-100)=1/2*. After the first game, participants of this experiment can choose: either to play the second game or to stop gambling. This possibility of playing the second game is presented in a few different experimental contexts:

**Experiment 1:** After the first game, a participant is not informed about the output of his first game. "Would you like to play the second game?"

**Experiment 2:** After the first game, a participant is informed that he/she won and earned 200 USD. "Would you like to play the second game?"

**Experiment 3:** After the first game, a participant is informed that he did not win, so and he lost 100 USD. "Would you like to play the second game?"

Denote these experimental contexts as **C1, C2, C3.**

In the series of experimental studies in cognitive psychology(see, e.g., Busemeyer and Bruza (2012) or Khrennikov (2010)), it was shown that in context **C1** participants in general reject the second gamble, but they agree to continue in both contexts, **C2** and **C3**. (This is the probabilistic statement and its rigorous probabilistic formulation will be present later.) Such behavior contradicts to the axiom of rationality which is typically formulated in the form of the *Sure Thing Principle* (STP) formulated by Savage (1954) who illustrated it by the following behavioral example:

*"A businessman contemplates buying a certain piece of property. He considers the outcome of the next presidential election relevant. So, to clarify the matter to himself, he asks whether he would buy if he knew that the Democratic candidate were going to win, and decides that he would. Similarly, he considers whether he would buy if he knew that the Republican candidate were going to win,*

*and again finds that he would. Seeing that he would buy in either event, he decides that he should buy, even though he does not know which event obtains, or will obtain, as we would ordinarily say. It is all too seldom that a decision can be arrived at on the basis of this principle, but except possibly for the assumption of simple ordering, I know of no other extralogical principle governing decisions that finds such ready acceptance."*

The essence of violation of STP which was demonstrated in a series of experimental studies in cognitive psychology is that experimental contexts **C1, C2, C3** are *incompatible*. One cannot know and not know the output of the first game at the same time or win and lose at the same time.

In quantum mechanics the thesis about existence of incompatible experimental contexts was formulated by N. Bohr in the form of *the principle of complementarity*, see (Plotnitsky, 2006, 2009) for the detailed discussion. There exist incompatible experimental contexts, e.g., contexts for measurements of system's position and momentum or the projections of spin or polarization onto different axes.

It is interesting that N. Bohr borrowed his principle of complementarity from psychology, from reading of James (1890). W. James (1890) considered the most fundamental complementarity in processing of information by the brain, namely, *complementarity of unconscious and conscious presentations* of information. It is clear that the unconscious and conscious presentations are incompatible. Thus nowadays by applying the methodology of quantum mechanics to cognitive science and psychology, we just reapply the basic principle elaborated by W. James.

However, nowadays this principle is applied in the novel mathematical formulation based on representation of systems' states in *the complex Hilbert state space W* and observables by *Hermitian operators* acting in *W*. Since operators can be noncommutative, the observables represented by such operators are constrained by *Heisenberg uncertainty relations* (in the general form given by the Schrödinger inequality). This constraint is interpreted as preventing the joint high precision measurement of two incompatible quantum observables (represented by noncommuting Hermitian operators). In the orthodox Copenhagen interpretation the Heisenberg uncertainty relation has even the stronger interpretation: two incompatible quantities cannot even be assigned jointly to a quantum system.

Thus by applying the mathematical formalism of quantum mechanics to biology, i.e., by proceeding in the framework of the quantum bioinformatics, we explore the possibility to model incompatible performances of biological systems: from the gene expressions to mental performance by human beings.

In quantum mechanics the principle of complementarity is typically connected with the existence of the *fundamental quantum of action* given by Planck's constant *h*. This was the original interpretation due to N. Bohr. However, in quantum bioinformatics we cannot refer to such a fundamental quantum of action. Here even the notion of the "biological energy'' and, hence, the notion of the "biological action" are not well defined, see, however, Khrennikov (2010) for attempts to define properly the "mental energy" (a form of the biological energy).

### 6.3. Contextuality in quantum physics and biology

As we have seen in analysis of the disjunction effect, incompatibility is the straightforward consequence of *contextuality*. Here contextuality is treated very broadly as dependence of outputs of measurements (expressions, performances, decision making, and judgement) on contexts. In fact, N. Bohr also pointed to the fundamental role of such broadly defined contextuality in quantum mechanics. At a few occasions he stressed that the whole experimental arrangement has to be taken into account. One of the main contributions of Khrennikov (2010) to quantum bioinformatics (starting with the works on quantum-like modeling of cognition, Khrennikov (1999, 2004a) was the understanding that the contextuality is the genuine source of incompatibility-complementarity.

Once again, we recall that here contextuality is considered in the very broad sense. We remark that in interpretational discussions in quantum physics (especially about violation of the Bell inequality), contextuality is reduced to contextuality of the joint measurement of two observables. In such a framework the measurement of one observable, say *B,* is considered as a part of context for the measurement of another observable, say *A*.

In a series of works (Khrennikov, 2001, 2003, 2004a,b,c, 2005) culminating in the monographs (Khrennikov, 2004, 2009, 2010), the second coauthor developed *contextual probability theory*. This theory is a natural extension of the classical probability theory (Kolmogorov, 1933) based on a single probability measure *p*. In contextual probability theory, a family of probability measures (with corresponding algebras of events) is explored to represent a multi-contextual group of measurements. The probabilistic structure of quantum mechanics, *"quantum probability"*, can be treated as one of possible models of contextual probability theory.

Adaptiveness of biological systems to the surrounding environment is the source of fundamental contextuality of their behavior. To model such behavior, it is useful to apply models of contextual probability theory, in particular, quantum probability (and hence quantum information theory). Of course, the reference to biological contextuality does not justify the use of the concrete contextual probabilistic model - quantum probability, see Khrennikov et al. (2015b) for the discussion. However, pragmatically it is natural to proceed with quantum probability as the most well developed theoretical formalism of probabilistic modeling of contextuality which was successfully tested in numerous experimental studies.

## 6.4. Violation of the law of total probability by quantum physical and biological systems

We recall that in classical probability theory, the probability update is based on the *Bayes formula* defining conditional probability and its consequences, especially *the law of total probability*. In the case of two dichotomous random variables x and y taking the values x=x1, x2, and y=y1, y2, respectively, this formula has the form:

$$p(x=xj) = p(y=y1) \, p(x=xj| y=y1) + p(y=y2) \, p(x=xj| y=y2).$$

In quantum probability theory, the probability update is based on a different (non-Bayesian) update rule. In the simplest case this is the state update corresponding to the *projection postulate* (due to von Neumann and Luders). In particular, the quantum probability update violates the formula of total probability. This leads to novel rules for probability inference. Quantum probability inference is one of the most successfully applied non-Bayesian probability inferences.

We remark that the first mathematically rigorous demonstration of violation of the formula of total probability in quantum theory was presented in the paper of Khrennikov (2001). Then similar argument was presented in the general contextual probability theory (Khrennikov, 2003, 2004a,b,c, 2005 2009, 2010). Applications to cognition were considered in a series of works of Khrennikov (2004a, 2010). The first experimental violation of the formula of total probability in cognitive science was demonstrated by Conte et al. (2004, 2007) with the experimental design based on the paper (Khrennikov, 2004a,b ). The experiment of Conte et al. (2004) was based on *recognition of ambiguous figures* – the well-established domain of cognitive science. The corresponding quantum-like model for the experimental statistical data was the first attempt to explore the mathematical formalism of quantum mechanics to model the process of recognition (concretely of ambiguous figures). Studies in this direction, experimental and theoretical, were continued at Tokyo University of Science, see Asano et al. (2014), Accardi et al. (2016).

### 6.5. Entanglement and quantum nonlocality in physical and biological systems

Another distinguishing feature of the quantum formalism is *the state entanglement.* Finding an adequate interpretation of entanglement is one of the most complicated problems of quantum foundations. The most common interpretation is that entanglement is an exhibition of *quantum nonlocality*. This is also a complicated interpretational issue. In this paper we have no possibility to go deeper in these interpretational problems. We proceed with the operational approach to entanglement and quantum nonlocality.

If a compound system $S=S1+S2$ is in some entangled state $\phi$ (for simplicity, suppose that this is a pure state). Then any operation (in particular, measurement) on $Sj, j=1,2,$ modifies the state $\phi$ of the whole system $S$. Such modifications are not arbitrary; each operation of Sj modifies the state of S in the very special way. Such consistent nonlocal modifications are basic for quantum computing and some quantum cryptographic schemes. Entanglement is the main source of the speed up of computations performed by a quantum computer.

In physics nonlocality is a mystical feature of quantum systems - *spooky action at a distance.* It seems that in biology we can proceed with a simpler interpretation. In physics the main problem is that spooky action at a distance is practically instantaneous. In any event it propagates essentially quicker than the light velocity. And this super-luminary propagation is the problem of physical theory. Biological systems are not of such huge size (physical experiments on entanglement were performed at distances of 100-200 kilometers). In a biological organism "action at a distance" can be generated by signaling based on chemical or electromagnetic signaling, e.g., signaling between cells.

Thus we have enlightened the two basic features of the quantum formalism playing the crucial role in its applications both to physics and biology:

A). Encoding of *complementary features of a system* in the same state, pure (given by a normalized vector $\phi$ from the complex Hilbert state space *W*) or mixed (given by a density operator $\rho$ acting in *W*).

B). Existence of entangled states representing "nonlocal operations" on the state of a compound system *S=S1+…+Sn* generated by local operation on its subsystems *Sj*.

### 6.6. Quantum-like modeling of epigenetic plasticity

Recent epigenetic studies (especially in microbiology) demonstrated that non-genetic variation arising during the life cycle of a biological system can be transferred to offspring, see, e.g., (Jablonka and Raz, 2009). Such a process is known as *epigenetic inheritance*.

An environment can induce modification of the structure of epigenome, including DNA methylation and histone. Moreover, in some contexts these modifications can be inherited by the progenitors. This is the adaptive mutation. It can be, in principle, be treated as a kind of *neo-Lamarckian process.*, see (Jablonka and Raz, 2009) and Asano et al. (2015a) for discussions on this very complicated issue.

In such studies the term *epimutation* is used for a heritable change in gene's expression that does not affect the actual base pair sequence of DNA (Kohler and Grossniklaus, 2002).

*Cellular epigenetic inheritance* is a narrower aspect of epigenetic inheritance as discussed in the broad sense. It refers to epigenetic transmission in sexual or asexual cell lineages, and the unit of this transmission is the cell. We point to the main types of cellular epigenetic inheritance (CEI): the CEI based on self-sustaining regulatory loops, the CEI based on three-dimensional templating, the chromatin-marking CEI, and the RNA-mediated CEI. The concrete structure of different mechanisms realizing these CEIs is not completely clear (Kohler and Grossniklaus, 2002), (Jablonka and Raz, 2009).

However, all these CEIs are parts of one universal phenomenon: development of special adaptive features in the process of interaction with the environment with following transmission of these features from a mother cell to the daughter cells. It is promising to develop universal model of CEI describing all its types by using the same formalism. In future such an operational model can be completed by creation of detailed models for each CEI and their interrelations.

In (Asano et al. 2013, 2015a) there was presented an operational quantum-like model of CEI which is applicable to all its possible types. Our model is based on quantum adaptive dynamics which is mathematically realized with the aid of theory of open quantum systems. Our model, although it does not describe explicitly processes in cells and epigenomes, can be useful for molecular biology. It presents a general mathematical structure of CEI; it justifies the epigenetic inheritance as an adaptive dynamical process. Hence, by ignoring the details of cellular mechanisms we acquire knowledge on universal information processes beyond CEI.

In the quantum-like model of epigenetic inheritance, the state of epigenome is represented as an entangled state. In our model the genome is treated as a compound quantum-like system, $G=g_1+\ldots+g_n$, where $g_j$, $j=1,2,\ldots,n$, denote concrete genes. For each gene $g_j$, we consider the space of all its possible epigenetic mutations $W_j$. (We recall that each mutation modifies the expression of this gene.) The epigenetic state of $g_j$ is represented by superposition of all possible epimutations. (In this model, we consider only epimutations, i.e., we ignore the genuine gene mutations.) If epimutations for different genes were independent, the state of g would be presented as the tensor product of epigenetic states of separate genes. However, we would like to explore entanglement expressing mathematically nonlocal coupling between epigenetic states of genes in the genome G. Thus local changes of chromatin marks (in some gene $g_j$) induce the consistent change of chromatin marks for all genes. Similarly to quantum computing this speed up exponentially the process of epimutations and at the same time provides consistency of epimutations for different genes.

By using entangled states genome can rapidly adapt its epigenetic counterpart to the impact of the surrounding environment. This gives the possibility to generate the new essentially modified epigenome during one generation. As was already pointed out, *this is a genuine Lamarckian process.*[3]

In the quantum bioinformatics framework, the stabilization of the epigenetic state to the steady state (corresponding to fixation of epigenetic mutations throughout the genome) is modelled with the aid of the quantum master equation. The Markov approximation in the form of the GKSL equation is explored, see (Asano et al., 2013, 2015a) for details.

---

[3] In the Darwinian fashion, epimutations would be generated randomly in various parts of the epigenome. Then only the offspring carrying epimutations consistent with the environment would survive. However, in experimental studies of CEI it was clearly demonstrated that this is not the case.

## Conclusion

We analyzed the most common mechanisms of speciation and evolutionary innovation. The implementation of these mechanisms requires partial directivity of evolution to solve the problem of the origin of biological complexity. The algorithm of evolution includes the quantum processes of molecular interaction and detection of the environment. Experiments to test the proposed mechanism were considered.

## Appendix

### Formulation of the problem of classical optimal control theory

The following statement of the problem is the basis of the optimal control theory (see, for example, Pontryagin et al, 1986). Let the dynamics of the system be described by the following differential equation:

$$\dot{x}(t) = f(t, x(t), u(t))$$

where $f(t,x,u)$ – is the vector-function; $t$ – is time, $t \in T = [t_0, t_1]$ – is the interval of the system operation;

$x = (x_1, \ldots, x_n)$ – is the system state vector;

$u = (u_1, \ldots u_q) \in U$ – is the control vector, $U$ – is the set of acceptable control values. The initial time $t_0$ of the process is specified, whereas the time of the end $t_1$ is defined by the point in time when the particle reached a given surface for the first time.

The cost function (functional of control quality) is defined for the set of acceptable processes satisfying the equation:

$$I = \int_{t_0}^{t_1} f^0(t, x(t), u(t)) dt + F(t_1, x(t_1))$$

where $f^0(t, x(t), u(t))$ and $F(t_1, x(t_1))$ are continuous differentiable functions.

It is required to obtain the values of $x^*$, $u^*$, $t_1^*$ such that

$$I(x^*, u^*, t_1^*) = \min I$$

**Formulation of the problem of quantum optimal control theory**

According to (Rigatos, 2015), the main approaches to the control of quantum systems are: (i) open-loop control and (ii) measurement-based feedback control (see Wiseman and Milburn, 2010). In open-loop control, the control signal is obtained using prior knowledge about the quantum system dynamics and assuming a model that describes its evolution in time.

On the other hand, measurement-based quantum feedback control provides more robustness to noise and model uncertainty (Chen et al, 2009). In measurement-based quantum feedback control, the overall system dynamics are described by the estimation equation called the stochastic master equation or Belavkin's equation (Belavkin, 1983). An equivalent approach can be obtained using Lindblad's differential equation (Wiseman and Milburn, 2010).

This work was partially supported by the grant "Mathematical Modeling of Complex Hierarchic Systems" of Linnaeus University

It was also supported (A. Khrennikov) by the EU-project "Quantum Information Access and Retrieval Theory" (QUARTZ), Grant No. 721321

**Figure captions**

Fig.1. Movement towards the highest set of fitness

Fig.2. Algorithm of partially-directed evolution